# Mending Wall: On the Implementation of Censorship in India

Devashish Gosain[1], Anshika Agarwal[1], Sahil Shekhawat[1], H. B. Acharya[2], and S. Chakravarty[1]

[1] IIIT Delhi, New Delhi, India
{devashishg,ansika1448,sahil13083,sambuddho}@iiitd.ac.in,
[2] Rochester Institute of Information Technology, NY, USA
acharya@mail.rit.edu

**Abstract**

This paper presents a study of the Internet infrastructure in India from the point of view of censorship.

First, we show that the current state of affairs – where each ISP implements its own content filters (nominally as per a governmental blacklist) – results in dramatic differences in the censorship experienced by customers. In practice, a well-informed Indian citizen can escape censorship through a judicious choice of service provider.

We then consider the question of whether India might potentially follow the Chinese model and institute a single, government-controlled filter. This would not be difficult, as the Indian Internet is quite centralized already. A few "key" ASes ($\approx 1\%$ of Indian ASes) collectively intercept $\approx 95\%$ of paths to the censored sites we sample in our study, and *also* to all publicly-visible DNS servers. $5,000$ routers spanning these key ASes would suffice to carry out IP or DNS filtering for the entire country; $\approx 70\%$ of these routers belong to **only two private ISPs**. If the government is willing to employ more powerful measures, such as an IP Prefix Hijacking attack, *any one* of several key ASes can censor traffic for nearly all Indian users.

Finally, we demonstrate that such federated censorship by India would cause substantial *collateral damage* to non-Indian ASes whose traffic passes through Indian cyberspace (which do not legally come under Indian jurisdiction at all).

## 1 Introduction

The current study of Internet censorship is mostly focused on openly censorious countries – China [37,52,43], Iran [34], Pakistan [55], etc. Even world-wide studies of censorship [32] essentially focus on countries well known for their censorship. However, in practice, many other countries still implement some form of censorship, which may even be more insidious because citizens are barely aware of it (for example, Sweden [6] and France [4]). In this paper, we consider the case of India, a major emerging power with over 450 million Internet users [19]

(up from 180 million in 2013, and on track to overtake Europe, which has 520 million users in all). India has been ambivalent about its censorship policy for years [13] (for example, in August 2015, the government ordered 857 target sites blocked, then backtracked in the face of public outcry [24]), but in context of the fact that *legally* [3] the executive branch in India holds unqualified power to block information, it is natural to be concerned about free speech in India. We begin by asking what policy, and what mechanism the Indian government currently employs; how this might change in future; and what unintended effects such censorship might have on foreign traffic transiting Indian ASes.

Our first step was to formally approach the authorities, by filing a *Right to Information* [25] request (RTI), inquiring about the policies and mechanism the government uses to block content. While the policy itself was confidential, the government was willing to share that the responsibility for filtering lies with individual ISPs, and that they could implement any mechanism they choose[4], as long as they *uniformly comply* with the given censorship policy.

In practice, an ad hoc approach to filtering generally leads to inconsistencies and errors [54], especially during updates [48]. Our initial experiments suggest that this is indeed the case; filtering policies are highly inconsistent across ISPs (see table 1), contrary to the government's expectations as stated in the official response. The current "feudal" approach to policing the Internet in India, *viz.* allowing ISPs to implement their own censorship mechanisms (which, as we show, do not "strictly adhere" to government diktats), results in inconsistent censorship policy enforcement: for *e.g.*, our findings show that users may be able to evade censorship more easily when accessing pornographic sites via Airtel, a large private ISP that screens fewer sites, compared to others such as MTNL.

We next consider the question of how, in future, the government might enforce a unified censorship policy for the whole country. The usual mechanism to enforce a single policy, is to redirect all Internet traffic through a single point of control, where all the traffic can be monitored(this approach has been employed by Iran [34], Venezuela [7], and Saudi Arabia [60]). Even in the case of China, a whole layer of state-controlled ASes must be used to act as a filtering layer that provides Internet connectivity to other ASes [60]. Nearly all the filtering is carried out by two Autonomous Systems - AS 4134 and AS 4812 [62].

Can the government, in future, force all networks to re-route their traffic via a chosen ISP so as to monitor the network ? We note that India's Internet infrastructure was grown through a laissez-faire approach (closely correlated with the cellular networking boom), and now consists of $\approx$ 900 ASes (over 170 of which are ISPs) [28]; it would require a massive effort to redirect all traffic through this new provider. Quite likely, the amount of disruption caused by such a redirection would make it difficult for a democratic nation to implement by fiat.

---

[3] Information Technology Act of India 2008 (Section 69A)

[4] IP and URL blacklists [38] are common, but ISPs may choose to employ more invasive techniques, such as DNS Injection Attacks [47] or even IP Prefix Hijacking [35,46]

*Might the government implement filtering with the existing infrastructure, without necessarily enforcing traffic redirection?* For the existing network, is it possible to find a *small set* of "heavy-hitter" ASes (and network elements in these ASes) that can potentially monitor or censor traffic without too much collateral damage? More formally:

- *Is it feasible to filter/monitor India's Internet traffic? If so, how, and where?* Given that India has over 900 ASes,
    1. Are there a small number of key ASes and routers where the government can intercept most Indian traffic to censored sites?
    2. How does the number of censorious ASes required, vary with the censorship technique – *e.g.* IP blacklisting, DNS Injection, IP Prefix Hijacking?
- *How much collateral damage will traffic filtering cause?* Internet censorship by an "upstream" AS can lead to inadvertent traffic filtering for its customers. How much impact can Indian censorship have on traffic that simply transits Indian cyberspace?

To answer the above questions, in this paper, we map the AS-level paths from each Indian AS to the potentially censored websites (our test corpus includes not only the sites publicly announced as being blocked, but also others from public resources such as Herdict [12]). We then construct router-level maps within these ASes, using Rocketfuel [58]. Finally, we identify the "key" ASes and routers, *i.e.* those which appear in an overwhelming majority of paths (and which are, therefore, the logical locations for network filtering).

Our experimental findings reveal that ten ASes cumulatively intercept over 95% of the paths connecting Indian ASes to the sites in our study (i.e. potentially censored sites). Eight of these key ASes, acting together, can poison $\approx$ 99% of the network paths leading to DNS resolvers in India (as well as other publicly available services such as GoogleDNS and OpenDNS), thus censoring URL requests. Even more alarming, when we consider another mechanism of censorship - IP Prefix Hijacking - we find five ASes, each of which can individually poison the BGP routes for almost all ASes in the country. Even though the actual number of routers needed for such efforts varies dramatically (from 7 in some ASes, to as high as 1782), overall, a total of less than 5000 routers across all the eight ASes are required for IP or DNS filtering – about 70% of which routers belong to two large private ISPs and any one of five key ASes is enough, if the government resorts to more aggressive measures like IP Prefix Hijack.

Finally, we note that paths that transit Indian ASes but originate outside India form a substantial fraction of the Internet: if India were in fact to adopt a comprehensive censorship scheme in its key ASes, she would censor about 1.15% of *all* Internet paths to the censored sites, worldwide.

Thus, the above findings would indicate that, in fact, ordinary Indian citizens **should** be concerned about censorship, and perhaps start to equip themselves with anti-censorship tools [39].

We begin by discussing the background and related work, in the next section.

## 2 Background and Related Work

The interaction of the Internet with government policy (especially censorship and privacy issues) is a controversial subject [15,30,14]. Our case study in this paper, India, is a democratic nation, but there is sufficient evidence of Indian censorship [8,21] that anti-censorship research organizations declare India "partly-free" [20]. For example, the Indian government officially demands that organizations (*e.g.* Google Inc., Microsoft *etc.*) censor pages deemed objectionable [9].

At present, the government delegates the censorship of traffic to ISPs, as per ambiguous blacklists[5]. This loose approach to censoring traffic leads to inconsistent filtering across ISPs – some users may be able to evade censorship by virtue of their provider ISP.

The question arises whether the Indian government *can* impose a centralized filter (as seen in *e.g.* Iran). Creating a new AS and redirecting through it would have high costs in network disruption, latency, service quality, and so on. But such a process will not be necessary *if* the current structure of Indian Internet is already well suited for monitoring and censorship.

To determine the set of ASes and routers where adversary may install infrastructure for censoring large fraction of network paths, as they exist today, we generated AS and router-level maps of India. We used such maps to identify such key ASes and routers, and the impact they have.

### 2.1 Background

Our paper relies heavily on mapping the structure of the Internet, an area of research called *network cartography* [44]. The Internet consists of routers and hosts, but also has some further structure: the routers and hosts belong to Autonomous Systems, which are independent networks (independent in the sense, they themselves choose who to exchange traffic with). Consequently, Internet mapping proceeds at two levels:

1. *AS-level mapping.* For our research, we required Internet maps representing paths connecting IP address of censored site to various ASes. We thus chose Gao *et al.*'s [56] AS path mapping approach. Their technique uses publicly-available BGP routes (obtained from various Internet Exchange Points across the globe [31])) and the relationships between the ASes [41], and outputs a directed graph of the Internet connecting IP prefixes to all ASes of the world. Other AS-level mapping approaches, such as the CAIDA Ark Project [3] and *iPlane* [53], involve `traceroute` probes from various vantage points to IPs in different ASes. Such approaches rely on `traceroute` and are generally limited by the network locations and availability of the volunteered probing nodes; they may not provide the AS-level path between any two randomly chosen ASes.
2. *Router-level mapping.* An AS is not a black box, but contains hosts and routers. Mahajan *et al.* [58] show how the internal structure of an AS can be mapped, by a combination of `traceroute` probes, IP alias resolution[6], and reverse DNS lookups.

---
[5] Several authors have mentioned how these blacklists vary over time [11,1]

*Powers of the Adversary:* Our adversary is a censorious government. The adversary aims to filter Internet traffic, and for this purpose may perform IP filtering, DNS injection/URL Filtering, and IP prefix hijacking attacks. We note that even a government has limitations; for example, it would prefer to implement filtering at a small number of locations, rather than at *every* ISP network in the nation, because of both various political and technical factors (*e.g.* if changing the blacklist implies wide scale router level re-configuration, there will almost certainly be inconsistencies and failures in enforcement).

## 2.2 Related Research

Much of the study of modern Internet censorship was developed in the context of China [63,49,62,61], particularly the different censorship techniques employed and the network destinations filtered. For *e.g.*, Winter *et al.* [61] examine how the Chinese authorities use DPI-capable routers to detect Tor Bridges. Others, such as [33], explored the mechanics of DNS filtering and how China is contributing to collateral damage. A major step forward was made by Verkamp *et al.* [32], who deployed clients in 11 countries (including India) to identify their network censorship activities – IP and URL filtering, keyword filtering and DNS censorship *etc.* Later authors – Nabi *et al.* [55] in Pakistan, and Halderman *et al.* [34] in Iran – demonstrate different methods of censorship employed by their respective regimes, as well as different forms of content blocked. Such studies of censorship in repressive regimes are often limiting, as they require Internet access from almost all network locations inside the country (Nabi *et al.* were able to get access from only five locations, and Halderman from only one).

We take a different direction with this paper. While we begin by examining instances of network censorship in our target country (India), our main aim is to determine the *potential* for censorship, in case the regime decides to become more censorious. Specifically, how bottlenecked is the Indian Internet? Is it possible for the adversary to place censors in a relatively small set of ASes and routers, and still filter a large fraction of network paths (and thus potentially users)? - if so, this presents a much lower barrier to entry than monitoring in every AS.

The most relevant related work we are aware of, is Singh *et al.*'s study of how Internet censorship correlates to network cartography [59]. The authors show a strong correspondence between the *Freedom House Index* [5] of a nation and its Internet topology, and indeed, claim that a nation's network topology is the best indicator of a countrys level of freedom. Our work makes use of network topology as well: we use it to determine the "key" network locations (ASes and routers) where the adversary (censorious government) would rationally deploy censorship infrastructure, if its aim was to censor all or almost all Internet traffic in the country, and the impact of such measures on network paths originating both within and outside the nation (but transiting Indian ASes). We perform this study for various traffic filtering techniques in the following section.

---

[6] Different interfaces of the same router, with different IPs, are called IP aliases

# 3 Motivation, Problem Description and Methodology

## 3.1 Preliminary Findings and Motivation

Well-studied censorious countries, such as China, Iran, and Saudi Arabia, tend to have a very clear censorship policy. In contrast, India has a rather ad hoc approach: the government expects all ISPs to (independently) enforce its policies. We find that in practice, traffic filtering is *highly* inconsistent across popular Indian ISPs – the set of blacklisted sites varies by orders of magnitude.

| ISP | Website Categories | | | | | | | |
|---|---|---|---|---|---|---|---|---|
| | **Escort** (150) | **Music** (100) | **Porn** (50) | **Torrents** (30) | **Social** (20) | **Political** (20) | **Tools** (20) | **Misc.** (150) |
| Airtel | 50, 80, 20 | 82, 6, 12 | 1, 49, 0 | 13, 16, 1 | 8, 10, 2 | 2, 15, 3 | 1, 14, 5 | 80, 41, 29 |
| Vodafone | 24, 87, 39 | 95, 1, 4 | 2, 45, 3 | 16, 11, 3 | 8, 8, 4 | 0, 13, 7 | 4, 11, 5 | 70, 35, 45 |
| Sify | 12, 98, 40 | 1, 75, 24 | 1, 48, 1 | 6, 22, 2 | 0, 16, 4 | 0, 15, 5 | 1, 16, 3 | 11, 75, 64 |
| NKN | 11, 105, 34 | 57, 33, 10 | 1, 48, 1 | 10, 16, 4 | 4, 12, 4 | 2, 14, 4 | 1, 14, 5 | 65, 56, 29 |
| BSNL | 41, 69, 40 | 68, 12, 20 | 0, 45, 5 | 12, 14, 4 | 7, 10, 3 | 4, 12, 4 | 3, 14, 3 | 88, 27, 35 |
| MTNL | 27, 98, 25 | 81, 2, 17 | 45, 3, 2 | 15, 12, 3 | 9, 8, 3 | 14, 1, 5 | 2, 12, 6 | 73, 23, 54 |
| Siti | 23, 99, 28 | 28, 56, 16 | 44, 4, 2 | 14, 13, 3 | 9, 8, 3 | 1, 14, 5 | 1, 12, 7 | 86, 29, 35 |
| Reliance Jio | 0, 123, 27 | 0, 77, 23 | 0, 38, 12 | 2, 26, 2 | 0, 18, 2 | 0, 16, 4 | 0, 15, 5 | 0, 78, 72 |

**Table 1.** Censorship trends in India: Some initial results.

To study such inconsistencies, we selected a list of 540 potentially censored websites, divided into 8 different categories (ranging from escort services, to anti-censorship tools like *Tor* [40]). We then systematically observed the censorship policy in different ISPs, by trying to access our potentially-censored websites through them.

Table 1 summarizes our findings. The rows represent the ISPs, columns correspond to the category of site which being filtered, and each entry is a 3-tuple $(c_n, o_n, x_n)$ representing the number of each type of response – *censored*, *open*, and *inaccessible*. [7] For example, we probed 150 escort websites through the Airtel network, and observed 50 to be censored, 80 open, and 20 inaccessible.

---

[7] We explain these terms below.

- *Censored:* the ISP intercepted the requests, and responded with an HTML iframe displaying a filtering message (indicating that requested URL had been blocked as per the directions from the Department of Telecommunication).
- *Open:* Websites were accessible without filtering.
- *Inaccessible:* Websites were "down". There was not enough information to determine if the sites were inaccessible due to network or system outages, or requests were deliberately filtered or throttled by the ISP.

We note that the variation of censorship by ISP is quite dramatic: Airtel blocks only 1 out of the 50 pornographic sites probed, whereas MTNL blocks 45.

It is clearly difficult to get hundreds of independent ISPs to correctly comply with censorship orders. The question arises whether, *if* the government decides to enforce a single policy, it is able to do so. So the question arises, *are there a few key bottlenecks in the existing network, where filtering may be carried out*?

### 3.2 Problem Description

In our research we are particularly interested in finding a small set of key locations (ASes and routers) that intercept a large fraction of network paths. More specifically, our questions are as follows.

– Is it possible for the government to monitor/censor a large fraction of Internet traffic by controlling only a small number of network locations (*viz.* ASes and routers)?
– What fraction of traffic could be filtered, and who would be most affected?
– Would such censorship affect users outside the country as well?

### 3.3 Evaluation Methodology

*Identifying Potential Network locations for IP filtering:* In order to estimate the locations for installing IP filtering infrastructure, we built an AS-level map using paths in the Internet, then focused on Indian ASes and their connections. Our map was built using Gao's algorithm [?], which finds AS-level paths to the home AS of chosen IP prefixes (in our case, censored sites) from every other AS in the Internet. The algorithm uses links from known AS paths in BGP routing tables; we obtained tables from a number of vantage points [31].

Unlike other nations, which have an unambiguous list of blocked sites [55], India has no clear censorship policy. We created a corpus of sites blacklisted by various government decrees (as reported by popular media), and also added the sites reported as blocked in India by the crowd-sourced censorship-reporting sites like Herdict [12]. These included social media sites, political sites, sites related to unfriendly nations, and p2p file-sharing sites. Finally, we added to the list the adult sites popular in India (as per Alexa [2]).

We randomly sampled about 100 sites from this corpus. We then computed the paths between all Indian ASes and these prefixes. The ASes appearing in these paths were sorted by frequency of occurrence; we thus selected the few most frequent ones.

*Do these ASes appear in paths to other potentially blocked sites as well?* To answer such questions, we re-estimated our paths with another set of about 220 sites, chosen from the corpus. The heavy-hitter ASes for this new set of paths were the same as the ones found before.

*Intra-AS topology generation:* In the second round of experiments, we employed the Rocketfuel algorithm [58] to compute the router-level paths through 10 heavy-hitter ASes (i.e. major Indian ISPs), then identified the routers which occur in a large fraction of paths (i.e. the heavy-hitter routers in heavy-hitter ASes), as follows.

1. Using `planetlab` nodes, we ran `traceroute` probes to three representative IPs in each prefix advertised by the ASes and by their immediate (1-hop) customer ASes.
   `Traceroute` returned router level paths leading to and out of the said ASes.
2. From the `traceroute` trace, we chose the sub-paths consisting of router IPs advertized by the AS under study (*i.e.* router within the ASes, identified from [16]).
3. We resolved the aliases (corresponding to the discovered router IPs) with `Midar` [18] alias resolution tool.
4. Finally, from the discovered `traceroute` paths we selected the minimum number of routers which cumulatively intercept a large fraction of the paths. To do this we chose the following heuristic:
   - If total number of edge routers are less than total number of edge and core routers that intercept a large fraction of the paths (over 90%), then we selected the edge routers alone (as the set of edge routers cover 100% of paths through the AS).
   - Else, we selected the "heavy-hitter" (core plus edge routers), appearing in a very large fraction of the paths (over 90%); not all edge routers may appear as often as others (edge and core routers appearing in the discovered paths).

*Identifying Potential Sites for DNS based filtering:* Another common approach to censorship is to prevent the DNS service from resolving requests. The censor either instructs DNS servers (within its jurisdiction) to filter requests for blacklisted URLs, or installs infrastructure to intercept DNS queries on routers (en-route to DNS servers) and respond with bogus IPs or NXDOMAIN responses – also referred to as *DNS Injection* attack.

Filtering DNS requests, either by simply dropping them, or by responding with bogus responses, could be carried out at the DNS server. However, in a country like India, hosting more than 55000 DNS servers, distributed across different networks, reconfiguring *all* such servers to filter DNS queries for blacklisted sites would not be easy (besides simple disobedience, there would also be misconfiguration bugs, delays, and network downtime). It would be much more practical to identify a few ASes (and routers therein), that intercept all or almost all the network paths connecting DNS servers to all ASes in the country.

To identify key ASes for DNS injection, we began by identifying the DNS resolvers across all Indian prefixes. We probed IP prefixes of every Indian AS for available DNS servers (UDP port 53) using `nmap` [51], and noted whether the response was *open*, *filtered*, or *closed*. (*Closed* corresponds to ICMP `'destination`

`port unreachable`' message responses from the destination. *Open* means the client received a meaningful response. *Filtered* indicates that the client received no response [8].)

Each IP, for which we obtained a *filtered* or *open* response, was sent a request to resolve the IP address of some popular WWW destinations (*e.g.* `https://www.google.com`). Addresses that allowed resolution were added to our list of publicly available DNS resolvers.

Finally, using Gao's algorithm, we constructed a graph of prefix-to-AS paths connecting the IP prefixes corresponding to DNS resolvers, and all the Indian ASes. To find the ASes which would be most effective at DNS injection, we identified ASes at the intersection of a large number of these paths.

*Impact of IP Prefix Hijack Based Censorship:* In an IP Prefix Hijacking attack, malicious BGP routers advertise fake AS-level paths[9] in an attempt to poison routes to an IP prefix (see Figure 1), thus attracting a large volume of traffic [35,42,45,57,36].

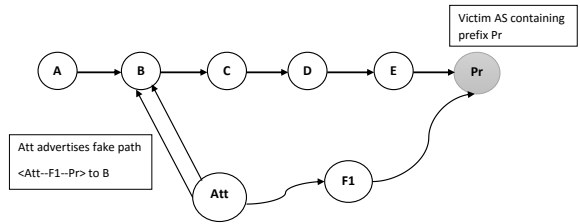

**Fig. 1.** IP Prefix Hijacking: Valid path: $A - B - C - D - E - Pr$. $A$ is the origin AS and $Pr$ the AS with the destination prefix. Attacker $Att$ advertises a shorter path $Att - F_1 - Pr$, to AS $B$. If $B$ chooses this path and directs its traffic to $Att$, the attacker can censor the traffic.

Prefix hijacking is an extremely aggressive attack, and unlikely to be used in practice; but it has been used in the wild (e.g. blocking of YouTube by Pakistani ISPs [23], and also those involving ConEd (US), TTNet (Turkey), Link Telekom (Russia) among others [46]) and remains viable as an orthogonal way of censoring traffic. So for completeness, we have also considered prefix hijacking as a potential tool for censoring the Internet in India.

In general, for a successful prefix hijack attack, the malicious AS either broadcasts a shorter path to the prefix, or claims to own it outright. The attacking AS advertises fake routes for the targeted prefix to all its neighbors. Ballani *et al.* [35] report that receiving ASes accept these advertisements based on the following heuristics:

---

[8] This may be due to unavailability or filtering by firewall(s)
[9] Alternatively, router misconfiguration can also lead to similar situations [54]

1. If there exists a customer path towards the target IP and iff the advertisement presents a shorter customer path, then choose it, else reject it.
2. If there exist a provider path towards the target IP and iff the advertisement presents a shorter provider path, then accept it. For all other cases, the paths are accepted without considering the length.
3. If there exist a peer path towards the target IP and iff the advertisement bears a shorter peer path, accept it. Customer paths are accepted without length considerations while provider paths are ignored.

**Estimating the Impact of Prefix Hijack Attack:** To study the potential impact IP prefix hijacking, we used the previously constructed AS-level topology and chose an attacker AS with a high *node degree*(i.e. the number of ASes adjacent to the said AS). Inspecting the prefix-to-AS paths, we identified ASes with which the attacker AS had a business relationship, and applied Ballani's heuristics to determine the number of ASes potentially affected by fake advertisements.

*Collateral Damage Due to Traffic Censorship:* Several non-Indian ASes rely on Indian ASes for Internet connectivity. Censorship activities in Indian ASes may potentially filter the traffic of these non-Indian customers as well [33]. For example, such unintended filtering was reported by Omantel, that peers with the Indian ISP Bharti Airtel [17]. As one of our research objectives, we try to identify ASes outside India that may be affected by Indian censorship. We identify paths which do not originate in India, but pass through or terminate in India. The non-Indian customers on such paths may face unwanted access restrictions.

## 4 Experimental Results

Continuing from the description of our experiment in the previous section, in this section we present our results. First, we consider router-level filtering, and how many ASes and routers must be selected for effective censorship (in terms of coverage of paths to filtered destinations). Along similar lines, we identify the locations where the adversary could launch a DNS injection attack. We go on to present the results of simulating IP prefix hijack attacks on Indian ASes. Finally, we report the collateral damage to foreign ASes due to IP filtering in India.

### 4.1 Network Locations for IP (Router-Level) Filtering

As mentioned earlier, we first obtained paths connecting Indian ASes to about 100 potential target sites (chosen from our corpus). Figure 3 represents the number of paths an individual AS intercepts; the horizontal axis of the graph indicates the ASes, ranked according to the number of paths each one intercepts. A small number of Indian ASes appear in the overwhelming majority of these paths; these ASNs and their owner organizations are presented in the table 4.1.

The question remains whether the ASes we observe are simply an artifact of the 100 target sites we chose. To check whether this is so, we repeated the experiment with another (non-overlapping) sample of 220 target sites from our corpus. The same 10 ASes covered the vast majority of paths to both sets of target sites, indicating that they are very likely major Indian providers of Internet infrastructure, and cover a majority of paths to *any* target sites.

| Rank | ASN | Owner |
|---|---|---|
| 1 | 9498 | Bharti Airtel |
| 2 | 4755 | Tata Comm. |
| 3 | 55410 | Vodafone |
| 4 | 9583 | Sify Ltd. |
| 5 | 9730 | Bharti Telesonic |
| 6 | 9885 | NKN Internet |
| 7 | 55824 | NKN Core |
| 8 | 45820 | Tata Teleservices |
| 9 | 18101 | Reliance Comm. |
| 10 | 10201 | Dishnet Wireless |

**Table 2.** AS Ranks, their ASNs and their owners.

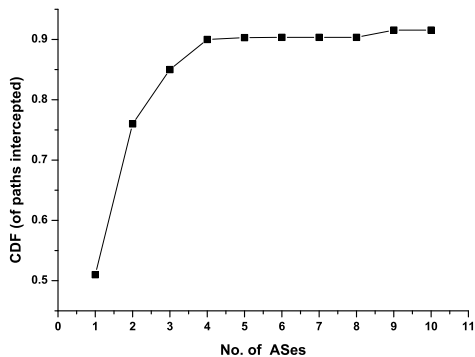

**Fig. 2.** CDF of Indian paths intercepted by ASes.

The cumulative results of *paths intercepted* vs *total number of ASes*, corresponding to both experiments, is presented in figure 2. As evident, *we only need 4 ASes to censor over* 90% *of the paths to the censored destinations, and* 10 *ASes for* 95% *of the paths.* Figure 3 represents the number of paths intercepted by each of these ASes individually.

*Intra-AS Topology:* We now consider the question of which *routers* (in our key ASes) are responsible for carrying the vast majority of Indian Internet traffic. Following Mahajan *et al.*'s approach [58] (as described previously in Subsec-tion 3.3), we create router-level maps of the key ASes, and identify routers that appear on a large fraction of the paths.

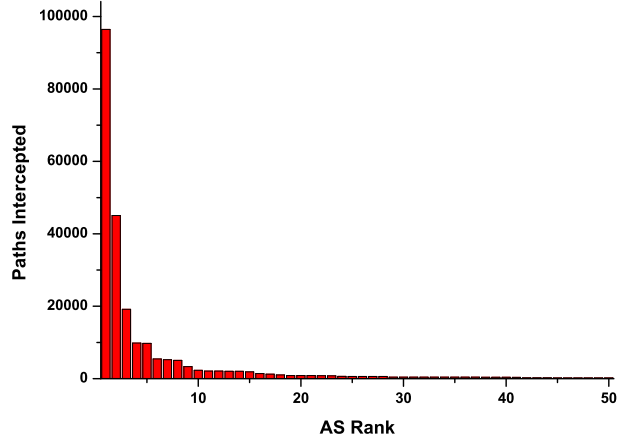

**Fig. 3.** Paths intercepted by individual ASes vs AS rank (by path freq.) Total 186679 paths from Indian ASes to 211 prefixes (hosting 320 potentially filtered sites).

Figure 4 shows the fraction of paths these routers cumulatively intercept. (For privacy concerns, we refrain from revealing the IP addresses of these routers.)

| ASN | # of Edge Routers (E) | # of Core Routers (C) | # of Heavy Hitter Routers (H) | # of DR's Required $min(E, H)$ |
|---|---|---|---|---|
| 9498 | 1782 | 5321 | 5192 | 1782 |
| 4755 | 1779 | 6229 | 6434 | 1779 |
| 55410 | 133 | 594 | 634 | 133 |
| 9583 | 484 | 4458 | 4275 | 484 |
| 9730 | 7 | 63 | 62 | 7 |
| 55824 | 66 | 325 | 254 | 66 |
| 45820 | 193 | 1147 | 1132 | 193 |
| 18101 | 462 | 2724 | 2677 | 462 |
| 10201 | 90 | 1396 | 1315 | 90 |

**Table 3.** The total number of edge and core routers in 9 ASes that appear in over 90% of the discovered paths. For *eg.,*. AS4755 has a total of 8404 routers (1779 edge + 6229 core). However, the total number of edge routers (1779) is less than the number of heavy hitters (6434).

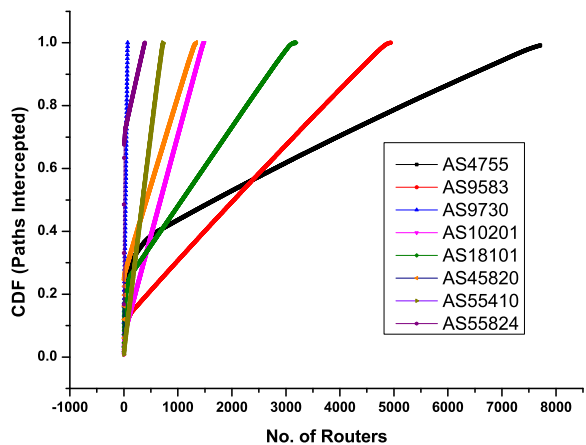

**Fig. 4.** CDF of `traceroute` paths intercepted by individual routers, sorted by increasing number of paths through each router (for 8 important ASes.)

Table 3 represents the number of edge and core routers that cumulatively appear in over 90% of the `traceroute` paths. The adversary could choose to place filters either at these points - heavy hitter routers of the heavy hitter ASes - or at the edge routers of the ASes, which together see all the traffic that passes through the AS. We find that the total number of edge routers is less than the number of "heavy-hitting" edge and core routers, and conclude that the lowest-cost solution for the adversary is to install censorship infrastructure on the (total of 4996) edge routers.

We note that, at present, the number of key routers varies significantly across ASes, from 7 to 1782. In case of the larger ASes, the AS network administrator could likely improve on our figures, by combining our findings with better information about the router-level topology and setting routing policy to pass all traffic through a smaller number of routers. Hence our count of 4996 routers is essentially an upper bound, limited by the policies of the present day.

*Collateral Damage:* Our graph of paths from censored prefixes to ASes has 186,679 paths of Indian origin (1.76% of paths). A comparable number - 121,931 paths of foreign origin (1.15% of paths) - transit through or terminate in an Indian AS. *Censorship by Indian ASes may inadvertently impact a very large number of unintended customers, across Finland, Hong Kong, Singapore, Malaysia, the US, and so on.*

### 4.2 Censorship Through DNS Filtering

Using our approach for identifying open DNS resolvers, we identified a total of 55,234 publicly accessible DNS servers from probing all 12.10 million Indian IPs.

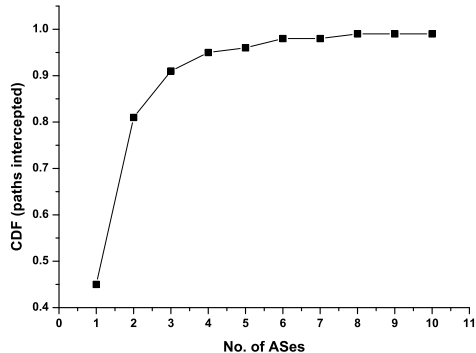

**Fig. 5.** CDF of DNS paths intercepted by top 10 Ases.

After identifying the prefixes corresponding to these each resolver IP, we selected one corresponding to each AS[10] In all, we selected 355 prefixes, representative of 355 unique Indian ASes. Finally, using Gao's algorithm, we estimated the paths from each Indian AS to the (prefixes corresponding to) DNS resolvers in India. *Cumulatively, 8 ASes (according to path frequency) can intercept* 99.14% *of these paths, and potentially launch DNS based filtering or Injection attacks (see figure 5).*

We note that these 8 ASes also appear among the 10 top ASes we identified for IP filtering and IP prefix hijacking. Hence, the same key routers for each of these ASes (as per Table 3) may be selected for installing infrastructure to launch DNS injection (or other DNS level filtering schemes). In all, 4906 routers across the 8 ASes can cumulatively filter DNS traffic for all Indian ASes [11].

### 4.3 Censorship Through IP Prefix Hijacking

For IP prefix hijacking, we chose to simulate attacks from the ASes with high node degree. Based on our censored-prefix-to-AS topology graph, we identified the top 10 ASes by node degree, and determined the number of ASes potentially vulnerable to attacks from each of these ASes. The results of these simulations are presented in table 4.

The table shows that a small number of ASes in India can potentially affect traffic from *all* Indian ASes, as well as a considerable number of foreign ones. For example, fake advertisements by *AS*4755 can impact a total of 955 ASes (896 Indian and 41 others). To effectively launch an IP prefix hijacking attack, the government needs control over the BGP speakers (which form a small fraction of all the routers of an AS); for ASes such as *AS*9730, with 7 edge and 63 core routers, this number is probably very small.

---

[10] For multiple prefixes belonging to same AS, we selected one with most resolvers.
[11] As mentioned in the previous sub-section, this number may be further reduced by routing optimization on the part of the AS network administrator.

| Owner Name | Attacking ASN | Number of Affected AS'es | |
|---|---|---|---|
| | | Indian | Non-Indian |
| Bharti Airtel Ltd. | 9498 | 896 | 59 |
| Tata Comm. | 4755 | 896 | 41 |
| Reliance Comm. Ltd. | 18101 | 896 | 41 |
| Vodafone Spacetel Ltd. | 55410 | 896 | 42 |
| Sify Ltd. | 9583 | 896 | 58 |
| Bharti Telesonic Ltd. | 9730 | 749 | 23 |
| Tata Teleservices | 45820 | 560 | 1 |
| Host Palace | 13329 | 896 | 45 |
| Dishnet Wireless Ltd. | 10201 | 896 | 24 |
| Idea Cellular Ltd. | 55644 | 896 | 37 |

**Table 4.** IP prefix hijack: A single AS (*e.g.* AS9498), is well capable of censoring the traffic of all 896 Indian ASes and few (59) non-Indian ASes through prefix hijack attack.

### 4.4 Analysis of Results

We observe that a very small number of ASes (less than 10) intercept a large fraction of AS-level paths connecting Indian ASes to our list of potentially censored sites (obtained from public announcements of censored sites in India), and that this affects a substantial number of foreign users as well. While this result is interesting, there remains the question of whether it applies to censored sites in general, or only the ones in our sample.

Our request to the Indian government, under its own *Right to Information Act* [25], for the complete list of censored sites [12], was refused by the Indian Government Department of Telecommunications and IT, citing confidentiality concerns. Therefore, to cross-validate our results we randomly sampled two sets of target sites from our corpus, and ran our algorithm on each in isolation. The same set of key ASes appeared in both sets. [13]

We believe that DNS filtering is a viable threat. Should the aforementioned ASes filter DNS requests, they would also impact over 99% of the AS-level paths connecting Indian ASes to DNS resolvers both within and outside India (particularly services such as GoogleDNS and OpenDNS). We note in passing that DNS filtering is more powerful than simple IP filtering: even if a censored site were hosted in a Content Distribution Network (CDN), a user would be unable to reach its content on the CDN, as the request would still have the URL of the origin site, and would thus be filtered.

Finally, while IP prefix hijacking is rarely used (owing to its potential to cause major network outages - *e.g.*, the Pakistan Government's blocking of Youtube [23]), there exist five Indian ASes, each of which could censor traffic for all (or nearly all) Indian users by launching an IP prefix hijack attack. Moreover, only a handful of routers in each of these ASes – *viz.* the BGP speaking routers may be sufficient for such attacks.

---

[12] RTI number: DOTEL/R/2017/50126

[13] We also note that these ASes are, in fact, partners to foreign network providers, and provide connectivity for almost every smaller AS in the country. This is perhaps unsurprising, given the hierarchical nature of the Internet as a whole [41].

## 5 Limitations and Future Work

### 5.1 Limitations

Our approach in this paper is to generate AS and router-level maps of India, and identify the key ASes and routers that intercept a large fraction of network paths. This approach is clearly limited to a snapshot of routing at a moment in time, and in fact we intend to see how our results vary over several years in future work. In addition, our AS-level and router-level mapping algorithms have the following limitations.

**AS path estimation (Gao's algorithm):** *Our path estimation strategy is limited by the quality of publicly-available BGP routes.*

- *Route-collector bias:* It has been argued by Gregori *et al.* that the existing route collectors (like routeviews [27], BGPmon [29], RIPE [26], PCH [22] *etc.*) miss many of the peering relationships between smaller ASes; our map, as it uses Routeviews data, inherits this weakness.
- *Incorrect route advertisements:* In general, BGP routes are known contain artifacts of misconfiguration and bogus advertisements [50,23]. Our estimated paths may also be contaminated with such artifacts.

**Router level topology estimation:** *The discovered topology may not reveal the actual router-level paths for packets traveling between the IPs of the probed AS and the censored websites.*

- *Router-level path variability:* Router-level maps of an AS are far more variable than AS-level maps: the latter rely on AS peering information (which is based on business relationships, that do not change frequently), while the former change with network conditions. Routing tables themselves are prone to inconsistencies and bogus routes [48,54].
- *Imperfect coverage by Traceroute:* We used a large number of `planetlab` nodes to launch `traceroute` probes [14], but there remains a chance that some routes are simply not covered; further increasing the number of vantage points, i.e. probing hosts, may improve our topology estimation by discovering new paths.
- *Routers filtering `traceroute` probes:* In many cases, routers are configured to not reply to `traceroute` probes with the usual `ICMP TTL Expired` messages, and remain anonymous, thereby reducing the accuracy of our estimated router-level topology.

---

[14] The `looking-glass` servers used by the original authors [58] were unavailable at the time of our experiments.

## 5.2 Future Work

Our study of Internet censorship in India can be directly extended to other nations; while our case study was done with Indian data, we make use of no features peculiar to India. We are currently extending our analysis to other countries, and developing metrics for how "centralized" a country is (i.e. how many key ASes it takes to censor traffic in a country), as well as how "central" it is in the global Internet (measured by the extent of collateral damage it can cause). There are several other directions to extend this research, which we will explore next.

First - objectionable content is frequently hosted on social media sites, or other sites with apparently benign URLs. Might the government target search engines and social networking sites as well(as seen in China)? Would this be a full blacklist, or partial? [15] And if so, would our key ASes be different for these target websites?

There is also the question of whether popular anti-censorship and anonymity preserving tools like Tor may be attacked by controlling a few network points. Finally, we also intend to consider the question of policing the cellular data network[16], in our future work.

## 6 Concluding Remarks.

Though the Indian state *declares* that it has a unified Internet censorship policy, the current state of censorship (where the responsibility of network filtering is left to individual ASes) is highly inconsistent. However, our results also show that *if* the Indian government wishes to impose a single policy, the structure of the Indian Internet shows that it would only need to control a small set of locations. (Furthermore, a significant fraction of network paths from foreign customers, which transit India, will be collateral damage for Indian censorship.)

1. Though India has $\approx$ 900 ASes, 10 ASes cover $\approx$ 95% of AS-level paths; a nationwide censor using IP-filtering functionality would need to control $\approx$ 5000 routers – a challenging, but tractable, number. In particular, two private ISP networks control over 70% of those routers (and may optimize the router selection further).
2. DNS based filtering requires only eight of these ASes and impacts $> 99\%$ of the AS-level paths connecting Indian ASes to the DNS resolvers both within and outside India (for services like GoogleDNS and OpenDNS).
3. Any one of five ASes is capable of disrupting network connectivity for all Indian ASes, through IP prefix hijacking attacks.

India, unlike China, is still ambivalent w.r.t. censorship, but the findings in this paper indicate that ordinary citizens should indeed be concerned (and possibly start to equip themselves with censorship circumvention techniques), as large scale censorship would not be very difficult for the government to implement.

---

[15] Semantics-based filtering is very hard; *e.g.* attempts to block jihadi mouthpiece sites also block sites that monitor jihad as a threat, such as `jihadwatch.org`.

[16] As per reports published in recent years, India has 860 million cellular users [10]